\documentstyle[prd,aps]{revtex}
\begin{document}

\draft
\twocolumn[\hsize\textwidth\columnwidth\hsize\csname
@twocolumnfalse\endcsname
\preprint{TIFR-TH/96-49, MIT-CTP-2565,}
\title{Universality of Low Energy Absorption Cross-Sections for Black Holes}
\author{Sumit R. Das$^1$,  Gary  Gibbons$^2$ and Samir D. Mathur$^3$}
\address{$^1$ Tata Institute of Fundamantal Research, Homi Bhabha Road, 
Mumbai 400 005, INDIA}
\address{$^2$DAMTP,  Cambridge University, Silver Street, Cambridge  
CB3 9EW,
United Kingdom}
\address{$^3$ Center for Theoretical Physics, 
Massachusetts Institute Of Technology, Cambridge, MA 02139
94305, USA}
\date{September, 1996}
\maketitle
\begin{abstract}
In this paper we compute the low energy absorption cross-section for
minimally coupled massles scalars and spin-$1/2$ particles, into a general
spherically symmetric black hole in arbitrary dimensions. The
scalars have a cross section equal to the area of the black hole,
while the spin-$1/2$ particles give the area measured in a flat 
spatial metric
conformally related to the true metric.

\end{abstract} 
\vskip2pc]

Recently there has been great interest in the possibility of relating
some of the properties of classical black hole solutions of the low
energy supergravity \cite{Nuff}
limits of string theories to a more fundamental
microscopic description based on strings and D-branes.  In particular
extremal black holes correspond in many cases to BPS states of the
theory, and their number for a given set of charges is expected to be
independent of coupling.  This allows a comparison between the number
of particle states (computable at weak coupling) with the Bekenstein-Hawking
entropy of the black hole (which would exist for the same charges at
strong coupling).  Agreement is found in all the cases investigated so
far \cite{extremal}
, thus suggesting that the Bekenstein-Hawking entropy for a hole does
indeed correspond to the count of microstates for the hole, though it
is still unclear where these microstates actually reside.

To study interesting processes like Hawking radiation, we need to allow
quanta to fall into the hole, rendering it non-BPS, after which it
would evaporate back towards extremality.  Are there relations between
the properties of particle states at weak coupling and properties of
black holes, when we consider deviations from extremality? One result
in this direction was presented in \cite{nonextremal} where it was shown that
if one naively ignores interaction between non-BPS states, then 
the degeneracy of a 
collection of branes and anti-branes continues to reproduce the
Bekenstein-Hawking entropy for non-extremal holes and leads to the
correct Hawking temperature. In \cite{dmone,dmtwo} it was
found that if one computes the low energy cross-section for absorption
and emission of neutral scalars in the 4+1 dimensional extremal black hole, 
then this cross
section agrees exactly with that for absorption or emission 
with the corresponding
collection of branes at weak coupling. This result has been
extended to charged scalars in 4 and 5 dimensions in  \cite{gubkleba}.
Recently it has been shown \cite{strommalda} that the D-brane
decay reproduces the correct grey body factors both for neutral
and charged scalar emission.

To discover if these are examples of a general pattern of universality
in the theory, we need to observe universalities that may exist in the
interactions of classical black holes. The absorption cross section
for low energy particles in 3+1 dimensional black holes was studied
extensively in the past, for example by Starobinski and Churilov
\cite{star}, Gibbons \cite{GG75} Page \cite{page} and Unruh
\cite{unruh}. In these calculations if we consider the particle to be
a massless minimally coupled scalar, then we find that the cross
section equals the area of the horizon of the black hole. In \cite{mw}
and \cite{dmone}, the 4+1 dimensional cases studied also yielded a
cross section equal to the area of the horizon. It was found in
\cite{kol}, however, that the low energy cross section for {\em fixed}
scalars (i.e. scalars which take fixed values at the horizon of some
extreme black holes) is suppressed by powers of the frequency.

In this paper we show that for all spherically
symmetric black holes the low energy cross section for massless
minimally coupled scalars is always the area of the horizon. Further,
we also find the corresponding result for minimally coupled massless
spin-$1/2$ quanta, and this also exhibits a universal form. Note that
the absorption processes studied here are not for black holes close to
extremality, though they do have the restriction to low energies.

We will consider general metrics in $(p+2)$ spacetime dimensions of the form
\begin{equation}
ds^2 = -f(r)dt^2 + g(r)[dr^2 + r^2 d\Omega_p^2]
\label{eq:one}
\end{equation}
where $d\Omega_p$ is the metric on the unit $p$-sphere.

At low energies only the mode with lowest angular momentum will
contribute to the cross section. For scalars this is the s-wave.
The mode $\phi_\omega (r)$ with frequency 
$\omega$ satisfies the equation
\begin{equation}
\{(r^p [f(r)]^{{1 \over 2}}[g(r)]^{(p-1)\over 2} \partial_r)^2 + 
\omega^2 (r^2 g(r))^p\}\phi_\omega (r) = 0
\label{eq:two}
\end{equation}
Define a coordinate $\tau$ by the relation
\begin{equation}
d\tau = {dr \over r^p [f(r)]^{{1 \over 2}}[g(r)]^{(p-1)\over 2}}
\label{eq:four}
\end{equation}
so that (\ref{eq:two}) becomes
\begin{equation}
[\partial_\tau^2 + \omega^2 (r^2 g(r))^p]\phi_\omega (\tau) = 0
\label{eq:five}
\end{equation}

Let the horizon be at the position $r = r_H$.  The area of the horizon is
\begin{equation}
A_H = r_H^2 g(r_H) \omega_p \equiv R_H^p \omega_p
\label{eq:seven}
\end{equation}
where $\omega_p$ is the volume of the unit $p$-sphere.

Close to the horizon we can write the solution of (\ref{eq:five})
by treating $r^2 g(r) \sim R_H^2$ to be a constant.
At the horizon 
  we want a purely ingoing wave, which is 
 given by $\phi_\omega (\tau) = e^{-i\omega R_H^p \tau}$.
At distances $r>>M$ but $r\omega<<1$ this solution behaves as
\begin{equation}
\phi_\omega \sim 1 - i \omega R_H^p \tau, ~~~~\tau \sim - 
{1 \over p-1} r^{-(p-1)}
\label{eq:fifteen}
\end{equation}

In the region $r\omega>>1$, the wave equation approximates to
\begin{equation}
[\partial_\rho^2 -{p(p-2) \over 4 \rho^2} + 1](\rho^{p/2}\phi_\omega
(\rho)) = 0
\label{eq:eleven}
\end{equation}
where we have used rescaled variables $\rho = \omega r$.
The solution is
\begin{equation}
\phi_\omega (\rho) = \rho^{(1-p)\over 2}[ A ~J_{({p-1 \over 2})}(\rho)
+ B~J_{-({p-1 \over 2})}(\rho)]    
\label{eq:fourteen}
\end{equation}
For $r\omega << 1$ this reduces to
\begin{equation}
\phi_\omega(\rho)~\sim {2^{-({p-1\over 2})} A\over \Gamma({p+1\over 2})} + 
{2^{({p-1\over 2})}\omega^{(1-p)}\over 
\Gamma({3-p\over 2})}~{B \over r^{p-1}}
\label{eq:fourteenp}
\end{equation}
whereas for $r\omega >> 1$ this becomes
\begin{eqnarray}
\phi_\omega~\sim \sqrt{{2\over \pi\rho^p}}
& &[A\cos(\rho-{\pi(p-1)\over 4}-{\pi\over 4})\nonumber \\
& & +B\cos(\rho+{\pi(p-1)\over 4}-{\pi\over 4})]
\label{eq:fourteenpp}
\end{eqnarray}
where for $p$ an odd integer, we take the analytic continuation
of all expressions in $p$, and the approximation
in the second part is valid for $r\omega<<1$. Matching onto 
(\ref{eq:fifteen}) we find
\begin{equation}
{B \over A} = i {2^{-(p-1)} (\omega R_H)^p \over p-1}
{\Gamma({3-p\over 2})\over \Gamma({p+1\over 2})} 
\label{eq:seventeen}
\end{equation}
which gives for the absorption probability of an $l=0$ 
 spherical wave
\begin{equation}
\Gamma=1-|{1+{B\over A}e^{i\pi (p-1)/2}\over 1+{B\over A}
e^{-i\pi (p-1)/2} }|^2
\label{eq:eighteen}
\end{equation}
\begin{equation}
=4{2^{-(p-1)}\over p-1} (\omega R_H)^p \sin[\pi(p-1)/2]
{\Gamma({3-p\over 2})\over \Gamma({1+p\over 2})}
\label{eq:eighteenp}
\end{equation}
in the limit $\omega\rightarrow 0$.
To convert the spherical wave  absorption
probability into
the absorption cross section we have to extract the ingoing
s-wave from the plane wave:
\begin{equation}
e^{ikz}\rightarrow e^{-ikr}r^{-p/2} Y_{00}K
\label{eq:twentyone}
\end{equation}
where $Y_{00}=\Omega_p^{-1/2}$ is the normalised s-wavefunction 
on the $p$-sphere. Here $\Omega_p = {2 \pi^{{p+1 \over 2}}\over
\Gamma({p+1 \over 2})}$ is the volume of a unit $p$-sphere.
This gives 
\begin{equation}
|K|^2={1\over 4\omega^p}
\Omega_p^{-1}\Omega_{p-1}^2 2^p(\Gamma[p/2])^2
\label{eq:twotwoi}
\end{equation}
so that the absorption cross-section $\sigma$ becomes
\begin{equation}
\sigma=\Gamma |K|^2={2\pi^{(p+1)/2} R_H^p\over \Gamma[(p+1)/2]}
=A_H
\label{eq:twentyfour}
\end{equation}
where $A_H$ is the area of the horizon. 

For minimally coupled massless spinors, the Dirac equation 
may be written down by making use of the properties of the
massless Dirac operator under conformal transformations
(see e.g. \cite{gsteif})
\begin{equation}
\nabla_\mu\gamma^\mu\psi=f^{-{1\over 2}}\gamma^0\partial_0[\psi]
+(fg^{p+2})^{-{1\over 4}}\gamma^i\partial_i[g^{{p\over 4}}
f^{{1\over 4}}\psi]=0
\label{eq:dione}
\end{equation}
Define $\chi=f^{1/4}g^{p/4}\psi$ and $h=\sqrt{f/g}$. Then the equation is
\begin{equation}
h\gamma^i\partial_i\chi=i\omega\gamma^0\chi
\label{eq:ditwo}
\end{equation}
Note that
\begin{equation}
\gamma^i\partial_i=\gamma^r[\partial_r+{p\over 2r}]+{1\over r}
(\gamma^i\nabla_i)_T
\label{eq:dithree}
\end{equation}
where the subsctipt $T$ stands for the part of the differential
operator tangent to the $p$-sphere. Write
\begin{equation}
\chi=\sum_{n = 0}^\infty F_n (r)\lambda^+_n + G_n (r)\lambda^-_n
\label{eq:difour}
\end{equation}
where $\lambda^\pm$ are mutually orthogonal
functions of the angular coordinates
only.  They satsify
\begin{eqnarray}
\gamma^r\gamma^0 \lambda^\pm_n & =& \lambda^\mp_n \nonumber \\
\gamma^r(\gamma^i \nabla_i)_T \lambda^\pm_n & = & 
\mp (n + {p \over 2}) \lambda^\pm_n
\label{eq:difoura}
\end{eqnarray}
Then we get
\begin{eqnarray}
h[\lambda^-_n(\partial_r+{p\over 2r})F_n & + &\lambda^+_n 
(\partial_r+{p\over 2r})G_n \nonumber \\
+{1\over r}(-(n+{p\over 2})F_n \lambda^-_n & + &(n+{p\over 2})G_n
\lambda^+_n)] \nonumber \\
& & =i\omega[F_n\lambda^+_n +G_n\lambda^-_n]
\label{eq:difive}
\end{eqnarray}
Setting to zero the coefficients of $\lambda^\pm$ we get
\begin{eqnarray}
h[\partial_r G_n +(p+n){G_n \over r}]& = &i\omega F_n \nonumber \\
h[\partial_r F_n -n{F_n\over r}]& = & i\omega G_n
\label{eq:disix}
\end{eqnarray}
The lowest angular momentum modes are found for $n=0$, which gives
(with $F \equiv F_0$)
\begin{equation}
\partial^2_r F +\partial_r F[{\partial_r h\over h}
+{p\over r}]+\omega^2h^{-2}F=0
\label{eq:diseven}
\end{equation}
Define the new coordinate $x$ through
\begin{equation}
{d\over dx}=(h(r)  r^{p}){d\over dr}
\label{eq:dieight}
\end{equation}

The equation becomes
\begin{equation}
\partial^2_x F+\omega^2 r^{2p} F=0
\label{eq:dinine}
\end{equation}
Again choosing an ingoing wave at the horizon, the analogue of
(\ref{eq:fifteen}) is
\begin{equation}
F=1-i\omega R_H^pg_H^{-(d-1)/2}{r^{-(p-1)}\over (p-1)}
\label{eq:diten}
\end{equation}
where $g_H = g(r_H)$ is the value of $g(r)$ at the horizon.
In (\ref{eq:diten}) we have used (\ref{eq:dieight}) to solve for $x$ for
large $r$
\begin{equation}
x={\rho^{-p+1}\over (-p+1)}
\label{eq:ditena}
\end{equation}
Comparing to the case of the scalar, we see that the absorption
probablity is $g_H^{-p/2}$ times the result for the scalar.

It is interesting to note for extremal holes $r_H \rightarrow 0$
and $g_H \rightarrow \infty$ so that the absorption cross-section
for minimally coupled fermions vanish in this limit of extremality.

The absorption probablility above implies a cross section
\begin{equation}
\sigma=2g_H^{-p/2}A_H
\label{eq:dieleven}
\end{equation}

Here the factor of $2$ comes from the two spinors
$\lambda^\pm$ that contribute to the absorption at low
energies, when the incident wave is a plane wave times
a constant spinor.

Note that (\ref{eq:dieleven}) is $2\omega_p r^p$, where $\omega_p r^p$
is the area of the horizon measured in the spatial metric
$ds^2=dr^2+r^2 d\Omega^2$, which is conformal to the spatial metric in
(\ref{eq:one}). Here $r$ is the isotropic radial coordinate.

It is well known that $N=2$ supergravity has
black hole solutions that preserve supersymmetry. In that case we
expect that the cross-section for the scalars and for their spinor
superpartners are related. The equation for these spinors is not
however the minimal Dirac equation, since the superpartners of
minimally coupled scalars have a coupling to the electromagnetic field
strength. This case is under investigation.

S.R.D. would like to thank Center for Theoretical Physics, M.I.T and
Physics Department, Washington University for hospitality. G.B. would
like to thank N. Straumann for hospitality and the Swiss National
Science Foundation for financial support at I.T.P, Zurich where part
of this work was carried out. S.D.M. is supported in part by
D.O.E. cooperative agreement DE-FC02-94ER40818.

\end{document}